\documentclass[%
 reprint,
 amsmath,amssymb,
 aps,
 prl,
]{revtex4-2}

\usepackage{graphicx}
\usepackage{dcolumn}
\usepackage{bm}
\usepackage[colorlinks=true,urlcolor=blue,citecolor=blue, linkcolor=blue]{hyperref}
\usepackage{siunitx} 
\usepackage{braket}
\usepackage{comment}
\usepackage{mathtools}
\usepackage{multirow}
\usepackage{lineno}

\begin{document}
\preprint{APS/123-QED}

\title{Observation of an Inner-Shell Orbital Clock Transition in Neutral Ytterbium Atoms}

\author{Taiki Ishiyama}
\author{Koki Ono}
 \email{koukiono3@yagura.scphys.kyoto-u.ac.jp}
\author{Tetsushi Takano}
\author{Ayaki Sunaga}
\author{Yoshiro Takahashi}
\affiliation{Department of Physics, Graduate School of Science, Kyoto University, Kyoto 606-8502, Japan}
\date{\today}


\begin{abstract}
We observe a weakly allowed optical transition of atomic ytterbium 
from the ground state to the metastable state $4f^{13}5d6s^2 \: (J=2)$ for all five bosonic and two fermionic isotopes with resolved Zeeman and hyperfine structures. 
This inner-shell orbital transition has been proposed as a new frequency standard as well as a quantum sensor for new physics. 
We find magic wavelengths through the measurement of the scalar and tensor polarizabilities and reveal that the measured trap lifetime in a three-dimensional optical lattice is 1.9(1)~s, which is crucial for precision measurements. We also determine the $g$ factor by an interleaved measurement, consistent with our relativistic atomic calculation.
This work opens the possibility of an optical lattice clock with improved stability and accuracy 
as well as novel approaches for physics beyond the standard model.
\end{abstract}
\maketitle

Recent development of optical atomic clocks using ions and neutral atoms has established a high fractional accuracy at the $10^{-18}$ level~\cite{Chou2010, Bloom2014, Ushijima2015}.
In addition to the contribution to metrology 
such as the redefinition of the second~\cite{Grebing2016, Milner2019} and geodesy~\cite{Takano2016, Lisdat2016}, 
a highly stable atomic clock enables various applications 
ranging from quantum simulations~\cite{Kato2016, Kolkowitz2017} 
to fundamental physics~\cite{SafronovaRMP2018} including gravitational wave detection~\cite{Kolkowitz2016}, and others. 
The development of even more precise clocks benefits all of these applications.

\begin{figure}[t]
\centering
\includegraphics[width = 0.95\linewidth]{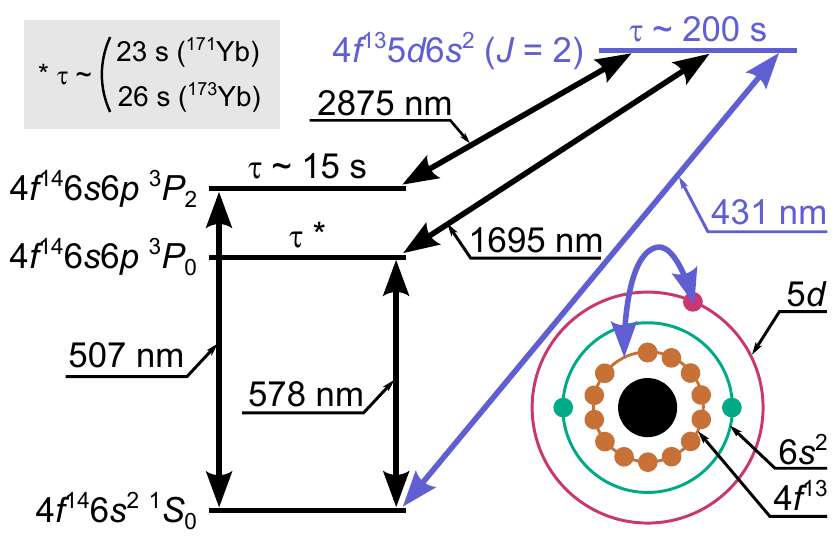}
\caption{\label{Energy diagram}
Clock transitions in an Yb atom. 
The relevant transition wavelengths and lifetimes $\tau$~\cite{Dzuba2018} are shown. 
}
\end{figure}

Recently, the optical transition of atomic ytterbium (Yb) to its metastable state $4f^{13}5d6s^2 \: (J=2)$ with an energy of 23188.518~cm$^{-1}$~\cite{NIST2022}, 
has been proposed as a new frequency standard with high stability and accuracy~\cite{Safronova2018, Dzuba2018}. 
As shown in Fig.~\ref{Energy diagram}, the radiative lifetime of the $4f^{13}5d6s^2 \: (J=2)$ state is calculated to be much longer than that of the other metastable states, which potentially improves the quality factor of an optical lattice clock.
In addition, the quadratic Zeeman shift as well as the black-body-radiation shift of the $^1S_0$ $\leftrightarrow$ $4f^{13}5d6s^2 \: (J=2)$ transition are calculated to be considerably small compared to those of the $^1S_0$ $\leftrightarrow$ $^3P_0$ and $^1S_0$ $\leftrightarrow$ $^3P_2$ transitions~\cite{Dzuba2018}, suggesting that the inner-shell orbital clock transition has a potential to reduce systematic uncertainties.
The dual clock operation~\cite{Akamatsu2018} combined with the well-established $^1S_0$ $\leftrightarrow$ $^3P_0$ transition, 
or already observed $^1S_0$ $\leftrightarrow$ $^3P_2$ transition~\cite{Yamaguchi2010} of Yb atoms, 
possibly can further reduce systematic uncertainties in a clock comparison. 

This metastable state of Yb also attracts considerable interest from the viewpoints of new physics searches~\cite
{Shaniv2018, Safronova2018, Dzuba2018}.
The $^3P_0$ $\leftrightarrow$ $4f^{13}5d6s^2 \: (J=2)$ as well as $^3P_2$ $\leftrightarrow$ $4f^{13}5d6s^2 \: (J=2)$~\cite{Tang2022} transitions show some of the highest sensitivities in neutral atoms to the variation of the fine-structure constant, 
providing also the possibilities of searching for ultralight scalar dark matter~\cite{Arvanitaki2015} and testing Einstein’s equivalence principle~\cite{SafronovaRMP2018}. 
The metastable state has also high sensitivity to the violation of local Lorentz invariance in the electron-photon sector~\cite{Kostelecky1999, Shaniv2018}, 
so far studied using single or two ions~\cite{Pruttivarasin2015, Megidish2019, Sanner2019, Dreissen2022} and dysprosium atoms~\cite{Hohensee2013}. 
Furthermore, establishing a new clock transition associated with the metastable state, in addition to the $^1S_0$ $\leftrightarrow$ $^3P_0$ and $^1S_0$ $\leftrightarrow$ $^3P_2$ transitions, 
provides a unique possibility to study a nonlinearity of the King plot using five bosonic isotopes of Yb 
in the search for a hypothetical particle mediating a force between electrons and neutrons beyond the Standard Model~\cite{Berengut2018, Ono2022}.
Clearly, the first step towards these experiments is to experimentally observe the associated transition and characterizations such as the lifetime~\cite{Safronova2018, Dzuba2018, Tang2022} and possible existence of a magic wavelength~\cite{Dzuba2018,Tang2022}.

In this Letter, we report the observation of the $^1S_0$ $\leftrightarrow$ $4f^{13}5d6s^2 \: (J=2)$ optical transition of Yb.  
The resonances for all five bosonic and two fermionic isotopes are clearly observed with resolved Zeeman and hyperfine structures. 
We find two magic wavelengths of 797.2(4)~nm and 834.2(1)~nm for the practical condition of $m_J$ = 0 and the laser polarization perpendicular to a magnetic field, through the measurement of the polarizability. 
Here $m_J$ is the projection of the total electronic angular momentum $J$ along the quantization axis.
In addition, our excitation and de-excitation sequence measurement of atoms in a three-dimensional optical lattice enables us to obtain the trap lifetime to be 1.9(1)~s.
Through an interleaved measurement with different magnetic fields, we determine the $g$ factor to be 1.463(4), consistent with our relativistic many-body calculation. 
These findings are promising for the development of an optical lattice clock with improved stability and accuracy, 
and its various applications for searching for physics beyond the Standard Model.



\paragraph*{\label{Experimental setup}Experimental setup.}
All spectroscopic measurements other than the lifetime measurement are done using ultracold Yb atoms evaporatively cooled in a crossed far-off resonance trap (FORT).
The number of atoms is about $2\times10^4 \sim 1.3\times10^5$ 
and the temperature is about $0.3 \sim \SI{1}{\mu \kelvin}$, depending on the isotope.
%
We apply a linearly polarized excitation light beam at a wavelength of 431~nm.
We measure the number of atoms remaining in the ground state after the irradiation of the excitation laser, and identify the optical resonance through the resonant atom loss. 
The angle of linear polarization, light intensity, and irradiation time are adjusted to optimize each spectroscopic measurement.
%
We note that the lifetime measurement is done for atoms loaded into a three-dimensional (3D) optical lattice.
See Sec.~S1 in Supplemental Material (SM) for the detail of the experimental system.

\begin{figure}
\includegraphics[width = 0.91\linewidth]{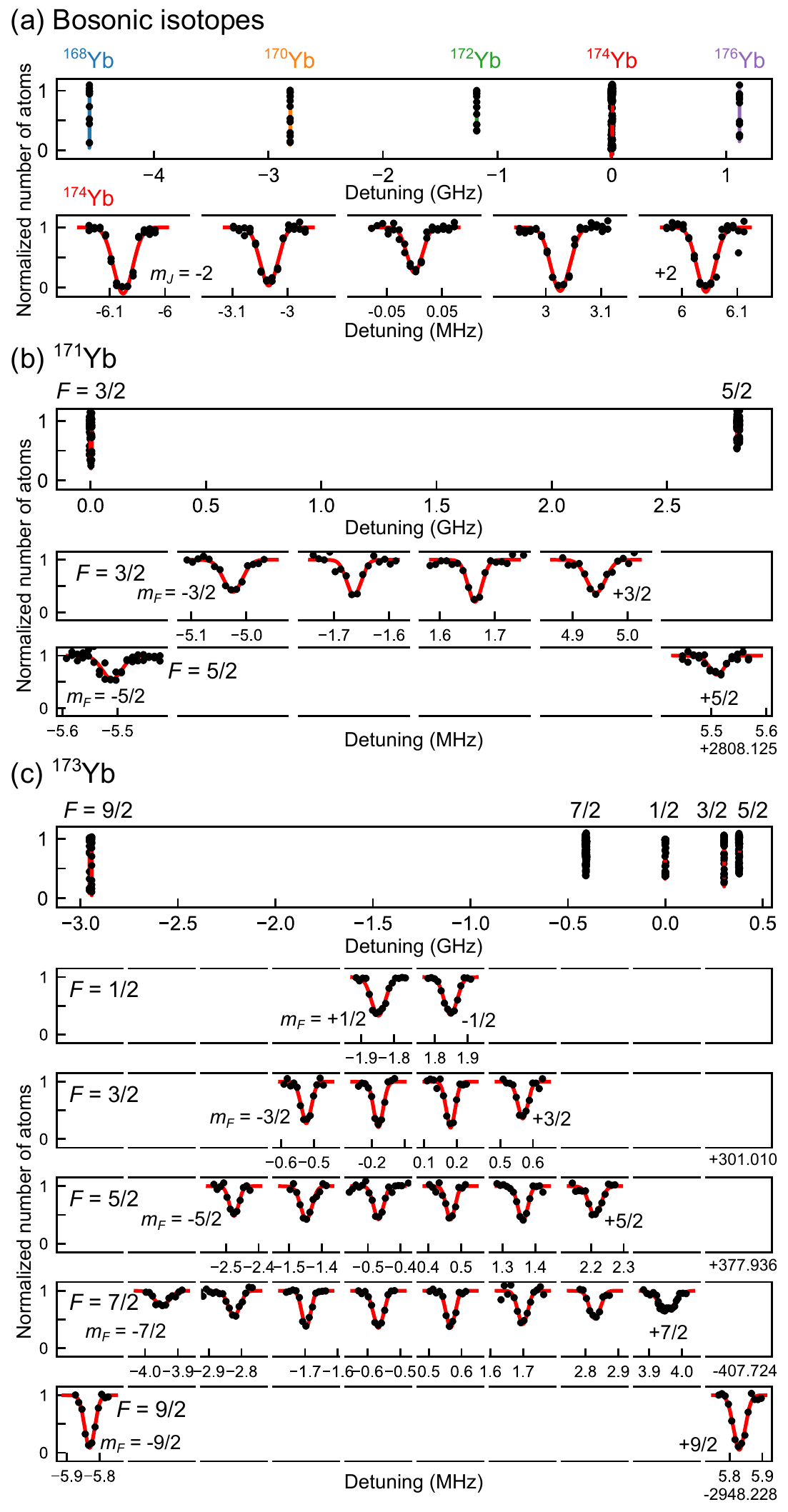}
\caption{\label{Excitation spectra}
Excitation spectra of all stable isotopes.
(a) Five bosonic isotopes.
The upper figure shows all spectra of the five isotopes for $m_J=0$ state, except for $^{174}$Yb, and the lower shows a magnified view around $^{174} \mbox{Yb}$.
The horizontal axis is the detuning from the resonance frequency of $^{174}$Yb ($m_J=0$).
(b, c) Two fermionic isotopes, (b) $^{171}$Yb($I=1/2$) and (c) $^{173}$Yb($I=5/2$).
The top figures show all spectra of all hyperfine states,
and the lower show $m_{F}$-resolved spectra of each hyperfine state. 
The horizontal axis is the detuning from the average of $m_F=\pm 1/2$ in $F=3/2$ for $^{171}$Yb and $F=1/2$ for $^{173}$Yb.
The solid curves are the fits with a Gaussian function.
Note that not all the magnetic sublevels for $F=5/2(9/2)$ state of $^{171}$Yb($^{173}$Yb) are shown.}
\end{figure} 

\paragraph*{\label{Zeeman and hyperfine spectra.}Zeeman and hyperfine spectra.}
Figure~\ref{Excitation spectra} summarizes the isotope shifts, Zeeman, and hyperfine spectra of the five bosonic isotopes ($^{168} \mbox{Yb}, ^{170} \mbox{Yb}, ^{172} \mbox{Yb}, ^{174} \mbox{Yb}$, and $^{176} \mbox{Yb}$ with nuclear spin $I=0$) and the two fermionic isotopes ($^{171} \mbox{Yb}$ and $^{173} \mbox{Yb}$, with $I=1/2$ and $5/2$, respectively).
The upper panel of Fig.~\ref{Excitation spectra}(a) shows the spectra of the five bosonic isotopes, and the lower panel shows $m_J$-resolved spectra of $^{174}$Yb.
The angle between the propagation (polarization) direction of the excitation light and the quantization axis, defined by a magnetic field, is 45$^\circ$(24$^\circ$).
This configuration allows the excitation to all magnetic sublevels of the $4f^{13}5d6s^2 \: (J = 2)$ state, according to the selection rule for a magnetic quadrupole (M2) transition.
%
The typical full width at half maximum (FWHM) is about 30~kHz, limited by the Doppler broadening of about 20~kHz and the finite laser linewidth of about 10~kHz. 
Note that more than half of the atoms are lost at the peak, suggesting some atom loss mechanism in the $4f^{13}5d6s^2 \: (J = 2)$ state, possibly caused by a shallow trapping potential and inelastic atom collisions.

Successful observation of the spectra of all bosonic isotopes enables us to determine the isotope shifts of the $^1S_0$ $\leftrightarrow$ $4f^{13}5d6s^2 \: (J = 2)$ transition.
See Sec.~S2 in SM for the detail of the measurement procedure.
The determined isotope shifts are summarized in Table~\ref{Isotope shift}, where only the statistical errors are evaluated.
Note that many systematic effects, such as the light shift and the quadratic Zeeman shift, are common among the isotopes within the accuracy of the present measurement.
In order to evaluate a possible line shift due to $s$-wave collisions between trapped ultracold atoms, which would be the largest systematic effect among many, we perform interleaved measurements with two different numbers of atoms by adjusting the evaporative cooling process. 
For the five bosonic isotopes, the resonance shifts, corrected for the difference in the atomic temperature, are within the uncertainties of about 5~kHz, and we find no evidence of atomic collision shifts at our measurement conditions.

\begingroup
\setlength{\tabcolsep}{10pt} 
\renewcommand{\arraystretch}{1.5} 
\begin{table}[tb]
\centering
\caption{\label{Isotope shift}%
Measured isotope shifts $\nu^{A'A}:=\nu^{A'}-\nu^{A}$ of the $4f^{14}6s^2 \: ^1S_0 \leftrightarrow 4f^{13}5d6s^2 \: (J = 2)$ transition, where $\nu^{A}$ is the transition frequency of an Yb isotope with a mass number $A$. Statistical 1$\sigma$ uncertainties are shown as $(\cdot)_\text{stat}$.}

\begin{tabular}{cr}\hline\hline
Isotope pair $(A',A)$ & Isotope shift $\nu^{A'A}$ $(\mathrm{MHz})$ \\
\hline
$(168,174)$ & $-4564.596(2)_{\text{stat}} $ \\ 
$(170,174)$ & $-2810.666(2)_{\text{stat}} $ \\
$(172,174)$ & $-1180.614(2)_{\text{stat}} $ \\
$(174,176)$ & $ -1115.766(6)_{\text{stat}} $ \\
\hline\hline
\end{tabular}
\end{table}
\endgroup

In addition, we precisely determine the $g$ factor of the $4f^{13}5d6s^2 \: (J = 2)$ state as $g_J = 1.463(4)$ with the uncertainty determined by the propagation of the fitting error and the uncertainty of the magnetic field calibration.
Our theoretical calculation of the $g$ factor using the DIRAC program~\cite{saue2020dirac,DIRAC22} gives $g_J = 1.465(2)$, which is consistent with the experiment.
See Sec.~S3 in SM for the detail of the measurement procedure and Sec.~S7 for the theoretical calculations.

Figures~\ref{Excitation spectra}(b) and (c) shows the hyperfine (top panel) and Zeeman spectra (lower panels) of $^{171} \mbox{Yb}$ and  $^{173} \mbox{Yb}$, respectively.
The observed Zeeman spectra for fermionic isotopes are well explained by the measured $g_J$.
Here, we identify an excitation mechanism other than the M2 transition.
A hyperfine interaction can cause the $4f^{13}5d6s^2 (J = 2)$ state to mix with other states such as $^3P_1$ and $4f^{13}5d6s^2 (J=1)$ states, making the electric-dipole (E1) transition partially allowed,
which is called a hyperfine-induced E1 transition~\cite{Boyd2007}.
The transition to the $F=3/2$ state of $^{171} \mbox{Yb}$  and those to the $F=3/2$, 5/2, and $7/2$ of $^{173} \mbox{Yb}$ are E1-allowed, and give resonance signals much stronger than other transitions that are allowed only by M2.

The data shown in Figs.~\ref{Excitation spectra}(b) and (c) enable us to determine the hyperfine constants of $^{171} \mbox{Yb}$ and $^{173} \mbox{Yb}$ of the $4f^{13}5d6s^2 \: (J = 2)$ state, respectively.
We obtain the magnetic dipole constant $A_\text{hfs}(^{171} \mbox{Yb})=1123.3(3)$~MHz, and $A_\text{hfs}(^{173}\mbox{Yb})=-309.46(4)$~MHz, and the electric quadrupole constant $B_\text{hfs}(^{173} \mbox{Yb})= -1700.6(9)$~MHz, with the uncertainties due to the fitting error. Note that the presumably much smaller systematic effects such as the collision shift are not evaluated at present. The careful evaluation of the systematic effects and precise determination of the hyperfine constants will be an important future work.
It is also noted that introducing a magnetic octupole constant $C_\text{hfs}$ for $^{173} \mbox{Yb}$ does not improve the fitting. 
See Sec.~S4 in SM for the detail.


\begin{figure}
\includegraphics[width = 0.95\linewidth]{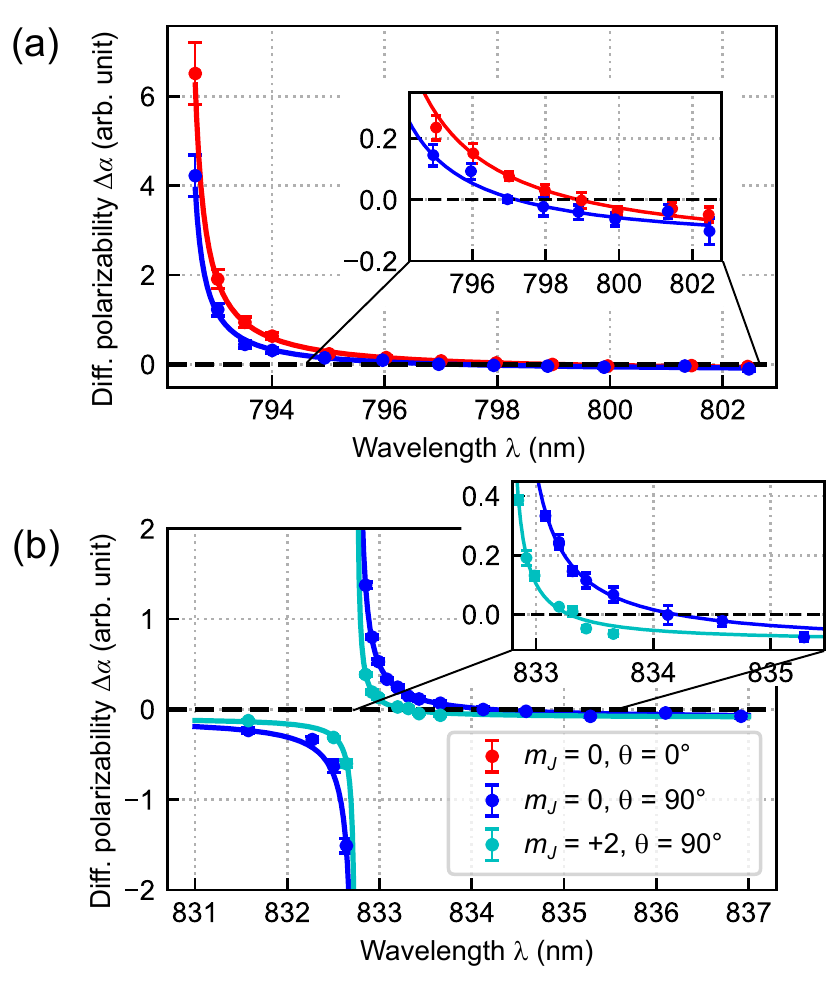}
\caption{\label{polarizability measurement}
Differential polarizability measurement between $^1S_0$ and $4f^{13}5d6s^2 \: (J = 2)$ states.
Measurements around the transitions: (a) $4f^{13}5d6s^2 \: (J = 2)$ $\leftrightarrow$  $4f^{13} 6s^2 6p_{3/2}\: (J=3)$ at $792.5$~nm, and (b) $4f^{13}5d6s^2 \: (J = 2)$ $\leftrightarrow$ $4f^{13} 6s^2 6p_{3/2} \:(J=2)$ at $833.7$~nm.
The differential polarizability, shown on the vertical axis as arbitrary units, 
is obtained from the linear fit to the measured differential light shifts $\Delta \nu$ as a function of the Ti:sapphire laser intensity.
The error bars on the vertical axis are obtained from the error propagation of 
the standard error of $\Delta \nu$ and the uncertainty of the Ti:sapphire laser intensity.
The solid curves are fits to the data.
}
\end{figure}

\paragraph{\label{Differential polarizability.}Differential polarizability.}
We perform a measurement of the wavelength-dependent polarizability of the $4f^{13}5d6s^2 \: (J = 2)$ state.
This is very important in the search for a magic wavelength at which the polarizabilities of the ground and  $4f^{13}5d6s^2 \: (J = 2)$ states coincide, and thus the line shift due to the trapping light is largely suppressed,
enabling ultra-narrow-linewidth spectroscopy and accurate clock operation~\cite{Katori2003}.
See Sec.~S5 in SM for the details of the experimental procedure.

In the case of linear polarization, the total polarizability $\alpha $ of the state $|J, m_{J}\rangle$ is
\begin{equation}
\alpha = \alpha^S  + \frac{3\mbox{cos}^2\theta -1}{2} \frac{3m_J^2 - J(J+1)}{J(2J-1)} \alpha^T,
\label{polarizability}
\end{equation}
where the superscripts $S$ and $T$ denote scalar and tensor, respectively, and $\theta$ is the polarization angle with respect to the quantization axis~\cite{LeKien2013}.
Using this formula, the differential polarizability $\Delta \alpha $ between the ground state with $J=0$ and the $4f^{13}5d6s^2 \: (J = 2)$ state is given in a straightforward manner.

In this work, we search for a magic wavelength within the tunable range of a titanium(Ti):sapphire laser which can provide enough power for optical lattices.
In particular, in this wavelength range, there are two E1-allowed optical transitions of $4f^{13}5d6s^2 \: (J = 2)$ $\leftrightarrow$  $4f^{13} 6s^2 6p_{3/2} \:(J=3)$ at a wavelength of 792.5~nm and $4f^{13}5d6s^2 \: (J = 2)$ $\leftrightarrow$ $4f^{13} 6s^2 6p_{3/2}\: (J=2)$ at 832.7~nm~\cite{NIST2022}, which will resonantly change the polarizability.
In fact, according to our calculation for the excited state and Ref.~\cite{Tang2018} for the ground state, the magic wavelengths are expected to be around 795.5~nm and 833.7~nm for $m_J=0$ and $\theta=90^\circ$.
See Sec.~S7 in SM for the theoretical calculation of the transition dipole moments for the two E1-allowed transitions.

As shown in Fig.~\ref{polarizability measurement}(a) and (b), we observe the resonant changes of the polarizability in the vicinity of the transition wavelengths of 792.5~nm and 832.7~nm. The solid curves are the fits to the data using the following equation: $\Delta \alpha (\omega) = a/(\omega_0^2-\omega^2)  + b$,
where $a$ and $b$ are fitting parameters, $\omega$ is the angular frequency of a Ti:sapphire laser, and $\omega_0$ is the resonance angular frequency of the $4f^{13}5d6s^2 \: (J = 2)$ $\leftrightarrow$ $4f^{13} 6s^2 6p_{3/2}\; (J=3)$ or $4f^{13}5d6s^2 \: (J = 2)$ $\leftrightarrow$ $4f^{13} 6s^2 6p_{3/2}\; (J=2)$ transition, given in Ref.~\cite{NIST2022}.
Note that this model assumes that only one resonant transition with $\omega_0$ dominantly contributes to the wavelength dependence, while the effects of other off-resonant transitions contribute as wavelength-independent offset constant $b$.
From the fitting, we can determine the magic wavelengths $\lambda_\text{magic}$ as follows:
\begin{equation}
    \lambda_\text{magic}=
    \begin{cases}
    \SI{798.9(4)}{nm} & (m_J=0, \;\theta=0^{\circ}),\\
    \begin{rcases*}
    \SI{797.2(4)}{nm}\\ \SI{834.2(1)}{nm}
    \end{rcases*}
    &(m_J=0, \;\theta=90^{\circ}),\\
    \SI{833.28(4)}{nm} & (m_J=2, \;\theta=90^{\circ}).
    \end{cases}
    \label{magic wavelength}
\end{equation}
Since two measurements with different $\theta$ or $m_J$ at each wavelength are performed, we can determine the differential scalar polarizability $\Delta \alpha ^S$ and the tensor polarizability $\alpha ^T $ from Eq.~\eqref{polarizability}.
Possible trap geometries to minimize the uncertainty of the tensor light shift on the performance of clock operation using the 797.2-nm optical lattice are discussed in Sec.~S5 in SM.
It should be noted that the differential higher-order light shift due to M1 and E2 multipolar polarizabilities and E1 hyperpolarizability~\cite{Porsev2018} should exist even at an E1 magic wavelength, where the differential light shift due to E1 polarizability vanishes~\cite{Katori2003}. The investigation of the operational magic condition for wavelength and intensity, where the overall light shift is insensitive to the lattice-intensity variation~\cite{Ushijima2018}, will be an important future work for clock operation.

\paragraph{\label{Lifetime}Lifetime.}
The lifetime of the excited state of a clock transition is important because it is directly related to the quality factor of the clock transition.
In this measurement, we use spin-polarized $^{173}\mbox{Yb}$ atoms in the $\ket{F=5/2, m_F=5/2}$ state loaded into a 3D optical lattice to minimize the possible atom loss due to inelastic interatomic collisions.
The 3D optical lattice consists of a 2D optical lattice at the wavelength $\lambda_1=759.4$~nm and a 1D optical lattice at the wavelength $\lambda_2=797.2$~nm, superimposed on the excitation light.
See Sec.~S6 in SM for the detail of the experimental sequence.
\begin{figure}
\includegraphics[width = 0.95\linewidth]{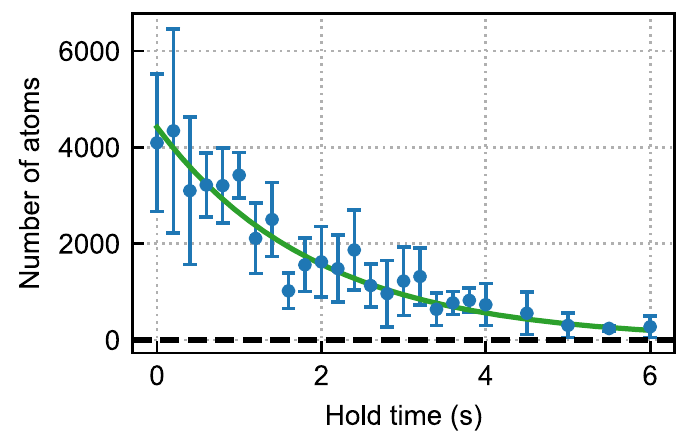}
\caption{\label{lifetime}
Lifetime measurement of atoms in $4f^{13}5d6s^2 \: (J = 2)$ state.
The horizontal axis shows the time to hold the excited atoms in the optical lattice, and the vertical axis shows the number of atoms in the excited state after the hold time.
The measurements are repeated four times, and the mean and standard deviation are plotted.
}
\end{figure}
Figure.~\ref{lifetime} shows the number of excited atoms as a function of the holding time in the optical lattice.
The lattice depths are 28$E_\text{R}$ for the 2D optical lattice at $\lambda_1$, and 25$E_\text{R}$ for the 1D optical lattice at $\lambda_2$. Here $E_\text{R} = h^2/(2m\lambda_2^2)=h\times1.8$~kHz represents the recoil energy at the wavelength $\lambda_2$, where $h$ is the Planck constant, and $m$ is the mass of an Yb atom.
From the fit to the data with a single exponential function, the trap lifetime of the excited atoms is obtained as 1.9(1)~s.
As a reference, we also investigate the trap loss of atoms in the ground state $^1S_0$ in the same experimental setup, and find the trap lifetime to be 2.8(2)~s, which would be limited by some mechanism such as collisions with background gases.
Assuming that the same mechanism also limits the trap lifetime of the $4f^{13}5d6s^2 \: (J = 2)$ state, we infer the intrinsic state lifetime of the excited state to be 5.9(1.3)~s.
This is compared with the theoretical calculations in Refs.~\cite{Safronova2018}, \cite{Dzuba2018}, and \cite{Tang2022}, which predict a lifetime of about 60, 200, and 190~s, respectively.
Our preliminary estimation shows that the rate of the photon scattering $\gamma_\text{sc}$ due to the lattice laser beam of near resonant 797.2-nm light at 25$E_\text{R}$ is $\gamma_\text{sc}=0.15$~s$^{-1}$, which seemingly explains the measured result.
Further systematic measurements with various experimental conditions will clarify the intrinsic lifetime of this state.

\paragraph{\label{Summary}Summary.}
We report the observation of the $^1S_0$ $\leftrightarrow$ $4f^{13}5d6s^2 (J = 2)$ optical transition of all seven Yb isotopes with resolved Zeeman and hyperfine structures. 
Two magic wavelengths of 797.2(4)~nm and 834.2(4)~nm for practical conditions are found and the measured lifetime is 1.9(1)~s, promising for applications for precision measurements. 
We also determine the $g$ factor to be $1.463(4)$, and compare the result with our relativistic many-body calculation with good agreement. 
A 3D optical lattice experiment with many ultracold atoms with isotope mixtures for the observed $4f^{13}5d6s^2 \:(J = 2)$ state is promising
to test the violation of local Lorentz invariance of electron-photon sector and Einstein’s weak equivalence principle.

\begin{acknowledgments}
This work was supported by the Grant-in-Aid for Scientific Research of JSPS (No. JP17H06138, No. JP18H05405, No. JP18H05228, No. JP22K20356), JST CREST (No. JP-MJCR1673), and MEXT Quantum Leap Flagship Program (MEXT Q-LEAP) Grant No. JPMXS0118069021, and JST Moonshot R\&D Grant No. JPMJMS2269. K. O. was supported by Graduate School of Science, Kyoto University under Ginpu Fund. A. S. acknowledges support from the JSPS KAKENHI Grant No. 21K14643. In this research work, A. S. used the computer resource offered under the category of General Projects by the Research Institute for Information Technology, Kyushu University.
\end{acknowledgments}

\providecommand{\noopsort}[1]{}\providecommand{\singleletter}[1]{#1}%

\end{document}


\renewcommand{\thefigure}{S\arabic{figure}}
\renewcommand{\thetable}{S\Roman{table}}
\renewcommand{\thesection}{S\arabic{section}}
\renewcommand{\theequation}{S\arabic{equation}}
\renewcommand{\bibname}{S\arabic{reference}}
\renewcommand{\bibnumfmt}[1]{[S#1]}
\renewcommand{\citenumfont}[1]{S#1}


\preprint{APS/123-QED}

\title{Supplemental Material for\\``Observation of an Inner-Shell Orbital Clock Transition in Neutral Ytterbium Atoms''}

\maketitle
\section{EXPERIMENTAL DETAIL}\label{sec:experiment}
\begin{figure}[t]
\includegraphics[width = 0.9\linewidth]{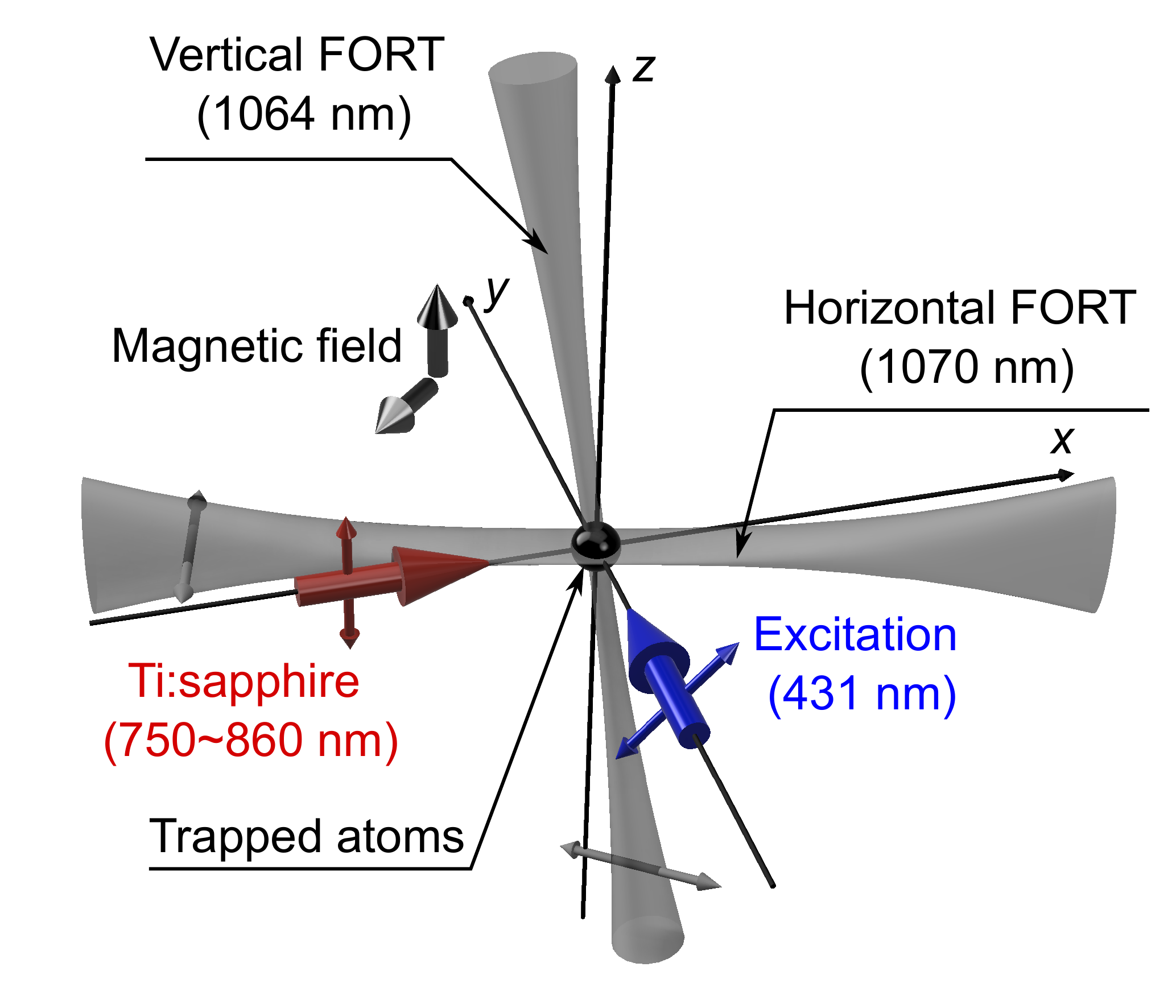}
\caption{\label{FigS1}
Schematic of our experimental setup. 
Our experiments are performed with Yb atoms in a crossed FORT consisting of 1064-nm and 1070-nm laser beams. 
The linearly polarized excitation light, shown as the blue arrow, is irradiated from the $y$-axis. 
For polarizability measurements, a linearly polarized Ti-sapphire laser, shown as the red arrow, shines along the $x$-axis. 
Note that the quantization axis defined by the magnetic field is changed for each measurement, as well as the polarization angles of the Ti:sapphire and excitation lasers.
}
\end{figure}

Figure~\ref{FigS1} shows the experimental configuration of our measurements. 
We note that the lifetime measurement is done for atoms loaded into a 3D optical lattice, although not shown in Fig.~\ref{FigS1}. 

For all isotopes, Yb atoms are decelerated by a Zeeman slower, collected in a magneto-optical trap using the $^1S_0$ $\leftrightarrow$ $^3P_1$ transition, and then evaporatively cooled in a crossed far off-resonance trap (FORT).
Since $^{171} \mbox{Yb}$($^{176} \mbox{Yb}$) has a small $s$-wave scattering length and thus evaporation cooling is inefficient, it is sympathetically cooled with $^{173}$Yb($^{174} \mbox{Yb}$)~\cite{Taie2010, Ono2022}.
The number of atoms and the temperature are measured by an absorption imaging method using the $^1S_0$ $\leftrightarrow$ $^1P_1$ transition.

The laser for the excitation at a wavelength of 431~nm is obtained by second harmonic generation of an interference-filter-stabilized external-cavity diode laser~\cite{Baillard2006} at a wavelength of 862~nm.
This 862-nm laser beam is offset-locked to an optical cavity made of ultra-low expansion glass with the Pound-Drever-Hall method~\cite{Drever1983}.
Note that the signal generator, driving a fiber electro-optic modulator for the offset lock, is referenced by the 10-MHz oscillator disciplined by the global positioning system.    
The output power of the 862-nm laser is about 30~mW, and then amplified to about 1~W by a tapered amplifier.
Using a periodically poled lithium niobate waveguide, the wavelength is converted to 431~nm and about 5~mW is obtained for irradiation of the atoms.
The peak intensity at the atomic position is up to about 0.5~kW/cm$^2$.


\section{ISOTOPE SHIFT}\label{sec:isotope}
As mentioned in the introduction, the isotope shift is an important physical quantity directly related to the search for new physics.
Here, we have measured the isotope shift of the transition to the $m_J=0$ state using five isotopes $^{168} \mbox{Yb}$, $^{170} \mbox{Yb}$, $^{172} \mbox{Yb}$, $^{174} \mbox{Yb}$, and $^{176} \mbox{Yb}$ with $^{174} \mbox{Yb}$ as a reference.
In order to minimize the systematic uncertainty of the isotope shift measurements, we use the following measurement technique:
after performing spectroscopy of one isotope, we re-lock the excitation laser frequency and perform the spectroscopy of the other isotope.
We repeat this sequence by multiple times.
By repeating the spectroscopy alternately, the effect of excitation light frequency drift during the measurement is minimized.
Note that the interleaved spectroscopy used in the polarizability and $g$ factor measurements is not applicable because the isotope shift is as large as several GHz, and it is difficult to change the laser frequency with the frequency-lock kept on.
When re-locking the excitation laser frequency, care should be taken to keep the offset voltage of the error signal and the magnitude of the transmitted signal of the optical resonator constant.
The FORT light intensity is also kept constant during the excitation to make the light shift due to the FORT light common in two isotopes.
Since the resonance shift is also atom-temperature dependent, the temperatures of the two isotopes are adjusted to be as close as possible.
The effect of residual temperature deviations is estimated to be smaller than 10~kHz.
The interaction shift due to collisions between atoms depends on the scattering length between atoms, which is isotope-dependent.
As is explained in the main text, no evidence of the collisional shift is found in our experimental condition.

\section{$g$ FACTOR}\label{sec:g-factor}
\begin{figure}
\includegraphics[width = 0.9\linewidth]{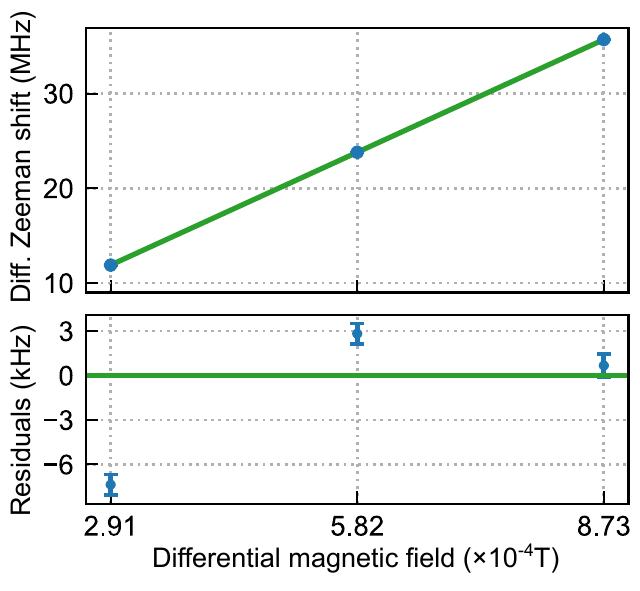}
\caption{\label{FigS2}
Measurement of the $g$ factor of the $4f^{13}5d6s^2 (J = 2)$ state.
In the upper panel, the differential Zeeman shifts of $^{174}$Yb ($m_{J}=+2$) are plotted as a function of the differential magnetic field, 
The solid green line is the linear fit to the data, the slope of which corresponds to the $g$ factor of the $4f^{13}5d6s^2 (J = 2)$ state.
The lower panel shows residuals from the linear fit.
The error bars on the vertical axis are the standard error.
}
\end{figure}

The measurement of the $g$ factor of the excited state $4f^{13}5d6s^2 \: (J = 2)$ is performed using the bosonic isotope of $^{174} \mbox{Yb}$.
The measurement sequence is as follows.
We apply a magnetic field along the $z$-axis to $^{174} \mbox{Yb}$ in a crossed FORT and excite the atoms to $m_{J}$ = +1 or +2.
Interleaved spectroscopy is performed to measure the Zeeman shift differences between four field strengths of 0.291(1), 0.582(2), 0.873(3), and 1.164(4)~mT, calibrated by spectroscopy with the $^1S_0$ $\leftrightarrow$ $^3P_1$ transition, in which the $g$ factor of the $^3P_1$ state $g_J(^3P_1)$ = 1.49282(2) is obtained from the weighted average of the two measured values in Refs.~\cite{Budick1967, Baumann1968}.
A linear fit to the magnetic field difference and the Zeeman shift difference, with the slope divided by the Bohr magneton $\mu_B$ and $|m_{J}|$, gives the excited state $g$ factor.
Note that the quadratic Zeeman shift is calculated to be -0.43~Hz/(mT$^2$) for $m_J=0$, and -0.34~Hz/(mT$^2$) for $\abs{m_J}=1,2$~\cite{Dzuba2018}, negligibly small compared to the uncertainty of our measurement.
Figure~\ref{FigS2} shows a typical result of such measurements for $m_J=+2$.
We perform the $m_{J} = +1$ and $+2$ measurements twice, and adopt the weighted average of the four measurements. The statistical uncertainty is obtained as the standard error of the four measurements, and the uncertainty of the magnetic field calibration, with a magnitude of about $0.3\%$, is considered as the systematic uncertainty.
In this way, we obtain the $g$ factor as $1.463(4)$, where the number in the parenthesis means the total uncertainty.


\begin{figure*}
\includegraphics[width = 0.9\linewidth]{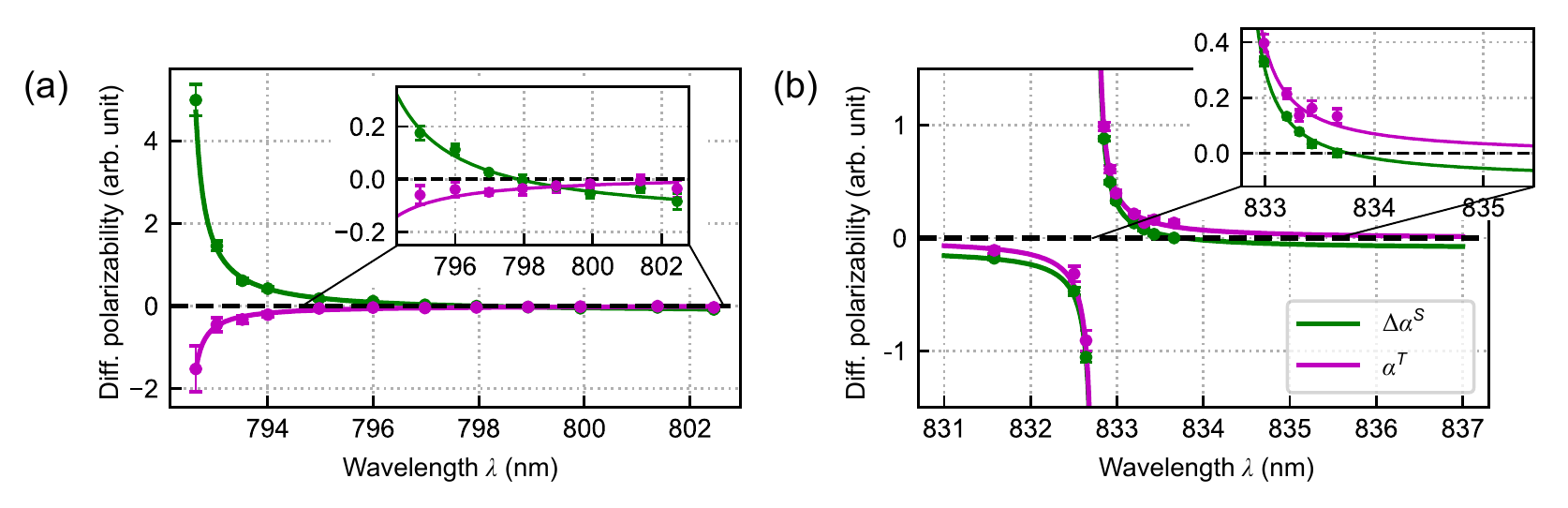}
\caption{\label{FigS3}
Differential scalar polarizability $\Delta\alpha^S$ and tensor polarizability $\alpha^T$ around the transitions: (a) $4f^{13}5d6s^2 \: (J = 2)$ $\leftrightarrow$  $4f^{13} 6s^2 6p_{3/2}\: (J=3)$ at $792.5$~nm, and (b) $4f^{13}5d6s^2 \: (J = 2)$ $\leftrightarrow$ $4f^{13} 6s^2 6p_{3/2} \:(J=2)$ at $832.7$~nm.
The solid curves are the scalar and tensor differential polarizabilities
expressed as the linear combinations of the curves in Figs.~3(a, b) of the main text using Eq.~(1).
For reference, the linear combination of the data points in Figs.~3(a, b) are shown.
We plot the average of the wavelengths of the two points for pairs which have slightly different wavelengths within $0.12$~nm.
}
\end{figure*}

\section{HYPERFINE CONSTANTS}\label{sec:hyperfine}
Here, we describe the details of the analysis to evaluate the hyperfine constants of the $4f^{13}5d6s^2 \: (J = 2)$ state for $^{171} \mbox{Yb}$ and  $^{173} \mbox{Yb}$. The hyperfine structure arises from the interaction between an electron with a total angular momentum $J$ and a nucleus with a spin $I$, and is expressed as a multipole series~\cite{Charles1955}. The hyperfine-structure Hamiltonian $\hat{H}_\text{hfs}$, truncated at the third term, is given by
\begin{widetext}
\begin{align}
    \hat{H}_\text{hfs} &= \hat{H}_\text{dip} + \hat{H}_\text{quad} + \hat{H}_\text{oct},
    \label{HFS}
    \\
    \hat{H}_\text{dip}  &= A_\text{hfs}\hat{\vb{I}}\cdot\hat{\vb{J}},
    \label{A}
    \\
    \hat{H}_\text{quad}  &= B_\text{hfs}\frac{3(\hat{\vb{I}}\cdot\hat{\vb{J}})^2+\frac{3}{2}(\hat{\vb{I}}\cdot\hat{\vb{J}})-I(I+1)J(J+1)}{2I(2I-1)J(2J-1)},
    \label{B}
    \\
    \hat{H}_\text{oct}  &= C_\text{hfs}\frac{10(\hat{\vb{I}}\cdot\hat{\vb{J}})^3+20(\hat{\vb{I}}\cdot\hat{\vb{J}})^2+2(\hat{\vb{I}}\cdot\hat{\vb{J}})[I(I+1)+J(J+1)+3-3I(I+1)J(J+1)]-5I(I+1)J(J+1)}{I(I-1)(2I-1)J(J-1)(2J-1)},
    \label{C}
\end{align}
\end{widetext}
where $\hbar\hat{\vb{J}}$ ($\hbar\hat{\vb{I}}$) is an operator for an electronic angular momentum (a nuclear spin). 
Eqs.~\eqref{A}, \eqref{B}, and \eqref{C} are associated with a magnetic dipole moment, an electric quadrupole moment, and a magnetic octupole moment, respectively. We consider an atom interacting with an external magnetic field $B$ along the $z$-axis, and the Zeeman Hamiltonian $\hat{H}_\text{Z}$ is given by
\begin{equation}
    \hat{H}_\text{Z} = \mu_\text{B}(g_J\hat{J}_z+g_I\hat{I}_z)B,
    \label{Zeeman}
\end{equation}
where $\hat{J}_z$ ($\hat{I}_z$) is the $z$-component of $\hat{\vb{J}}$ ($\hat{\vb{I}}$). As shown in Sec.~\ref{sec:g-factor}, the $g$ factor of the $4f^{13}5d6s^2 \: (J = 2)$ state is measured to be $g_J=$ 1.463(4), and $g$ factor of a nuclear spin is represented as $g_I=\mu_I/(\mu_BI)$, where a nuclear magnetic moment $\mu_I$ in units of the nuclear magneton $\mu_\text{N}$ is given by 0.49367(1) for $^{171}$Yb, and -0.67989(3) for $^{173}$Yb~\cite{Olschewski1972ZP, Stone2005ADNDT}. 
Here we do not consider the correction to $g_\text{I}$ due to mixing with other states caused by the hyperfine interaction~\cite{Porsev2004, Boyd2007}, since the correction is not known and, in the first place, the contribution of $g_I$ to the determination of the hyperfine constants is negligibly small, considering the uncertainties of the obtained hyperfine constants.
To determine the hyperfine structure constants $A_\text{hfs}$, $B_\text{hfs}$, and $C_\text{hfs}$, the eigenenergies of $\hat{H}_\text{hfs} + \hat{H}_\text{Z}$ are used for the fits to the data shown in Fig.~2 in the main text. As a result, we obtain $A_\text{hfs}(^{171} \mbox{Yb})=1123.3(3)$~MHz, and $A_\text{hfs}(^{173}\mbox{Yb})=-309.46(4)$~MHz, and $B_\text{hfs}(^{173} \mbox{Yb})= -1700.6(9)$~MHz, with the uncertainties due to the fitting error. 

\section{POLARIZABILITY}\label{sec:scalartensor}
\subsection{Calculation of magic wavelength}
Here we estimate magic wavelengths for the $^1S_0\leftrightarrow 4f^{13}5d6s^2\;(J=2)$ transition. An atom interacting with a laser field with intensity $I_0$ has an energy shift, called the AC Stark shift $\Delta E = -\alpha I_0/(2\epsilon_0c)$, where $\epsilon_0$ is the permittivity in vacuum. The polarizability for an atom in a fine-structure level $J$ is given by Eq.~(1) in the main text when the laser field is linearly polarized.
The scalar and tensor possibilities are expressed as follows~\cite{LeKien2013}:
\begin{equation}
\begin{aligned}
\alpha_{J}^{S} &= \frac{1}{\sqrt{3(2J+1)}}\alpha_J^{(0)},\\
\alpha_{J}^{T} &= -\sqrt{\frac{2J(2J-1)}{3(J+1)(2J+1)(2J+3)}}\alpha_J^{(2)},
\end{aligned}
\end{equation}
where
\begin{equation}
\begin{aligned}
\alpha^{(K)}_J = &(-1)^{K+J+1}\sqrt{2K+1}
\sum_{i}
\begin{Bmatrix}
1 & K & 1 \\
J & J_i & J
\end{Bmatrix}
\nonumber\\
&\times\frac{|\langle J_i\|\hat{D}\|J\rangle|^2}{\hbar}
\left(\frac{1}{\omega_i-\omega}+\frac{(-1)^K}{\omega_i+\omega}\right).
\end{aligned}
\end{equation}
Here $\{\cdot\}$ represents the Wigner 6-$j$ symbol, and $\langle J_i\|\hat{D}\|J\rangle$ is the reduced transition dipole matrix element.
To calculate the polarizability of the excited state, we consider two $E1$-allowed optical transitions of $4f^{13}5d6s^2 \: (J = 2)$ $\leftrightarrow$  $4f^{13} 6s^2 6p_{3/2} \:(J=3)$ at a wavelength of 792.5~nm and $4f^{13}5d6s^2 \: (J = 2)$ $\leftrightarrow$ $4f^{13} 6s^2 6p_{3/2}\; (J=2)$ at 832.7~nm, whose transition wavelength are within a tunable range of a Ti:sapphire laser. The corresponding transition dipole moments are presented in Sec.~\ref{sec:results}. We calculate magic wavelengths, where the polarizability of the excited state coincides with that of the ground state $^1S_0$~\cite{Tang2018}. As a result, we obtain two magic wavelengths for $m_J = 0$ and $\theta=90^\circ$: $\lambda_\text{magic}$ = 795.5~nm and 833.7~nm. 
Similarly, the polarizability of an atom in a hyperfine-structure level $F$ is given by
\begin{equation}
\alpha = \alpha_F^S + \frac{3\cos^2\theta-1}{2}\frac{3m_F^2-F(F+1)}{F(2F-1)}\alpha_F^T,
\label{polarizability: hyperfine}
\end{equation}
where
\begin{align}
\alpha_F^S=&\alpha_J^S,\\
\alpha_F^T = &-(-1)^{J+I+F}\sqrt{\frac{2F(2F-1)(2F+1)}{3(F+1)(2F+3)}}
\nonumber\\
&\times
\begin{Bmatrix}
F & 2 & F \\
J & I & J
\end{Bmatrix}
\alpha_J^{(2)}.
\end{align}

As shown in Fig.~3 in the main text, we measured the total differential polarizability $\Delta\alpha$, and one can retrieve the differential scalar polarizability $\Delta\alpha^S$ and tensor polarizability $\alpha^T$ from the measured $\Delta\alpha$ using the following simultaneous equations:
\begin{equation}
    \begin{cases}
    \Delta\alpha(m_J=0,\theta=0)=\Delta\alpha^S-\alpha^T, \\
     \Delta\alpha(m_J=0,\theta=90^\circ)=\Delta\alpha^S+\frac{1}{2}\alpha^T,\\
     \Delta\alpha(m_J=2, \theta=90^\circ)=\Delta\alpha^S-\frac{1}{2}\alpha^T. 
    \end{cases}
    \label{simultaneous equations}
\end{equation}
Figure \ref{FigS3} shows the retrieved $\Delta\alpha^S$ and $\alpha^T$ from the measured $\Delta\alpha$ using Eq.~\ref{simultaneous equations}.

\subsection{Experimental procedure}
In our measurement, we apply a high magnetic field of 14.6~mT to $^{174} \mbox{Yb}$ in a crossed FORT and irradiate $\pi$-polarized excitation light at 431~nm to induce a magnetic-field-induced $E1$ transition~\cite{Taichenachev2006} to the $m_J = 0$ state in the $4f^{13}5d6s^2 \: (J = 2)$.
The wavelength is tunable between 750~nm and 860~nm, with a maximum intensity of 8~kW/cm$^2$ at the atom position.
The laser power is stabilized by feedback control using an acoustic-optic modulator, and the beam pointing is monitored to evaluate the uncertainty of the laser intensity at the atom position.
By interleaved measurements with three light intensities $0, I_0, 2I_0$ of the Ti:sapphire laser, we determine the differential light shift $\Delta \nu$ with respect to the light intensity difference.
This interleaved measurement minimizes systematic effects such as the drifts of the excitation light frequency and the Ti:sapphire light intensity.
From a linear fit of the differential light shift $\Delta \nu$ with a light intensity difference, we obtain the differential polarizability $\Delta \alpha$ as its slope.

\subsection{Possible trap geometries for optical atomic clock}
Different from the case of the $^1S_0$ $\leftrightarrow$ $^3P_0$ transition, the $^1S_0$ $\leftrightarrow$ $4f^{13}5d6s^2 \: (J = 2)$ transition is influenced by the tensor polarizability, which depends on $\theta$, and so fluctuations in the magnetic field or the polarization of the laser light will cause fluctuations of the light shift.
Since $d(\cos^2\theta)/d\theta=0$ at $\theta =0$ or $90^\circ$, $\theta =0$ or $90^\circ$ is preferable.
Here, we estimate how much this contributes to the future clock operation using the following trap geometries.
\begin{enumerate}
\item 
Let us consider a 3D optical lattice clock with a single occupancy~\cite{Akatsuka2008} using $^{174} \mbox{Yb}$ at the condition of $m_J=0$ and $\theta=90^{\circ}$, and the operation at a magic wavelength $797.2(4)$~nm with a depth of $25E_R$ for each lattice axis.
\item
Another case of interest is a 1D optical lattice clock~\cite{Takamoto2006} operated with excitation to the stretched state $\ket{F=5/2, m_F = \pm5/2}$ of spin-polarized $^{171}$Yb atoms, which is useful to reject the higher-order Zeeman effect induced by the hyperfine structure.
From the scalar and tensor polarizabilities, shown in Figs.~\ref{FigS3}(a) and Eq.~\ref{polarizability: hyperfine}, the magic wavelength for $^{171}$Yb is calculated as 797.3~nm at the condition of $\theta = 90^\circ$.
\item
In addition, single-atom arrays trapped by optical tweezers are a promising platform for an optical atomic clock~\cite{Young2020}. We consider tweezer-trapped $^{174}$Yb atoms at the condition of $m_J=0$ and $\theta= 90^{\circ}$, and the operation at a magic wavelength $797.2(4)$~nm with a depth of $25E_R$.
\end{enumerate}
For these three cases shown above, the fluctuation $\delta\theta$ of $\theta$ causing a fractional uncertainty at the $10^{-18}$ level is estimated to be about 1~mrad, which is manageable in the current experimental technique.

 \begin{figure}[t]
\includegraphics[width = 0.9\linewidth]{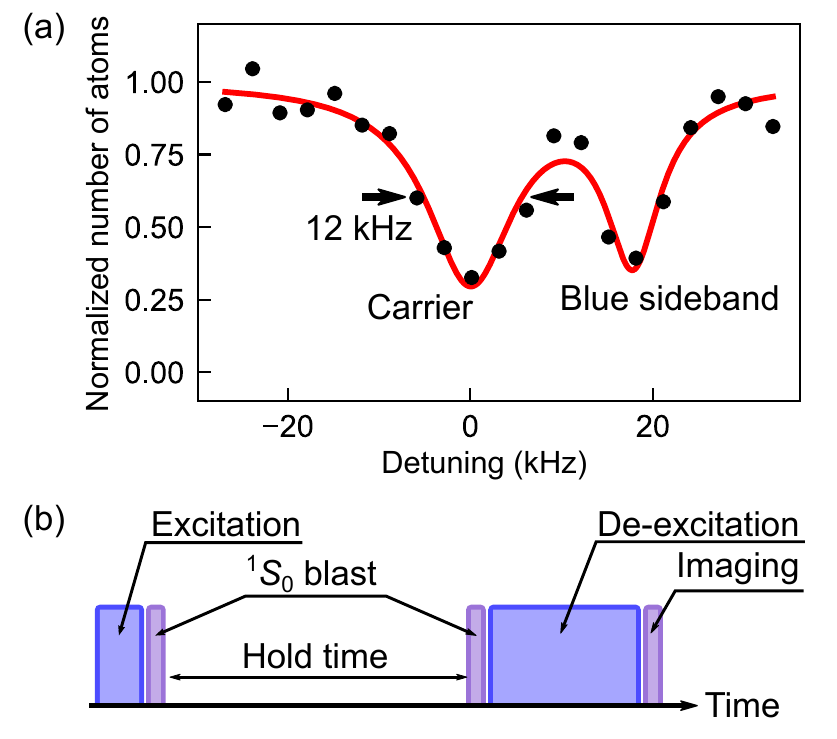}
\caption{\label{FigS4}
(a) Excitation spectrum of $^{174}$Yb ($m_J=0$) in a 3D optical lattice.
The horizontal axis shows the detuning of the clock laser, and the vertical axis shows the normalized number of atoms after the irradiation by the excitation laser.
The solid red curve is the fit with a double-peak Lorentzian function. The FWHM of the carrier spectrum is 12~kHz.
(b) Pulse sequence of the lifetime measurement.
The blue and purple pulses correspond to the $^1S_0\;\leftrightarrow\;4f^{13}5d6s^2\;(J=2)$, and $^1S_0\:\leftrightarrow\:^1P_1$ transitions, respectively.
First, the excitation pulse of 3 ms is applied. 
Then, the remaining $^1S_0$ state is blown up with 399-nm light resonant to the $^1S_0$ $\rightarrow$ $^1P_1$ transition.
Then, we wait for variable hold time.
At the end of hold time, again we blow off the atoms that have decayed from the excited state to $^1S_0$.
Finally, we count the atoms remaining in the excited state by absorption imaging using the $^1S_0$ $\leftrightarrow$ $^1P_1$ transition, by the returning the excited atoms to the ground state with the de-excitation pulse of 20~ms light pulse.
}
\end{figure}
%
\section{Lifetime}\label{sec:scalartensor}
The lifetime of atoms in the $4f^{13}5d6s^2 \: (J = 2)$ state is measured using a 3D optical lattice. Figure~\ref{FigS4}(a) shows the spectrum with resolved carrier and blue-sideband components.
The full width at half maximum (FWHM) of the carrier is about 12~kHz, narrower than the 30~kHz linewidth for atoms in a crossed FORT and
is limited by the linewidth of the excitation laser and the effect of the differential light shift at the non-magic wavelength of the 759-nm optical lattice light.

%
Figure~\ref{FigS4}(b) shows the pulse sequence of the lifetime measurement.
Spin-polarized $^{173} \mbox{Yb} \: (m_F=+5/2)$ atoms are loaded into a 3D optical lattice and then excited to the $\ket{F' = 5/2, m_{F'} = +5/2}$ state by 3 ms excitation light, where $F'$ is the total angular momentum in the excited state.
The excitation light is linearly polarized along the $z$-axis, enabling the hyperfine-induced $E1$ transition with $\Delta m_{F} = 0$.
The excitation and de-excitation sequence depicted in Fig.~\ref{FigS4}(b) enables the measurement of the lifetime.
We also measure the lifetime of the ground state $^1S_0$.
After loading the lattice, we wait for the hold time, and finally observe the number of atoms with absorption imaging using the $^1S_0$ $\leftrightarrow$ $^1P_1$ transition. 
The lattice depths are 28$E_R$ for the 2D optical lattice at 759.4~nm, and 25$E_R$ for the 1D optical lattice at 797.2~nm. 
The lattice depth is calibrated using a pulsed-lattice method applied to a $^{174}$Yb Bose-Einstein condensate.

We estimate the photon-scattering rate for the atoms in the $4f^{13}5d6s^2 \: (J = 2)$ state due to the 1D magic-wavelength optical lattice laser beam at 797.2~nm, which is close to the resonance wavelength of the $4f^{13}5d6s^2 \: (J = 2)$ $\leftrightarrow$  $4f^{13} 6s^2 6p_{3/2}\: (J'=3)$ transition at 792.5~nm. The scattering rate $\gamma_\text{sc}$ is given by
\begin{equation}
    \gamma_\text{sc}=\frac{3\pi c^2\Gamma^2}{2\hbar\omega_0^3}\frac{\beta^2}{\Delta^2}I_\text{lat},
\end{equation}
where $\Delta = \omega-\omega_0$ is the detuning of the lattice beam from the resonance, and $I_\text{lat}$ is the intensity of the lattice laser. The transition strength factor, or the square of the Clebsch-Gordan coefficient $\beta^2 = \abs{\bra{J'=3, m_{J'}=\pm1}\ket{J=2, m_J = 0;1,\pm1}}^2 = 0.4$ is considered, since the polarization of the lattice light is perpendicular to the quantization axis. The spontaneous decay rate $\Gamma$ is given by
\begin{equation}
    \Gamma = \frac{2J+1}{2J'+1}\frac{\omega_0^3}{3\pi\epsilon_0\hbar c^3}|\langle J'\|\hat{D}\|J\rangle|^2.
\end{equation}
As a result, the rate of the photon scattering is obtained as $\gamma_\text{sc}=0.15$~s$^{-1}$. In the case of another magic wavelength $\lambda_\text{magic}=834.2(1)$~nm, the photon-scattering rate at the same condition is calculated as $\gamma_\text{sc}=0.066$~s$^{-1}$.

\section{Electronic structure calculations}\label{sec:results}
All of the electronic-structure calculations were carried out using the DIRAC program package~\cite{saue2020dirac,DIRAC22} (git hash dbeb14f, d86857b, 721ac5b). 
The Dirac-Coulomb-Gaunt Hamiltonian was employed in all calculations. 
The dyall v3z basis set~\cite{Dyall_Gomes_2010} was employed in the uncontracted form. 
We employed the KR-CI module~\cite{Fleig2003JCP,Knecht2010JCP,Knecht_thesis,Denis2015NJP} for the relativistic configuration-interaction calculation.
The electrons in $5s$, $5p$, $4f$, and $6s$ orbitals were correlated in the calculations of the $g$ factor. The electrons in $5p$, $4f$, and $6s$ orbitals were correlated in the calculations to obtain the polarizabilities, 
that is, the $E1$ transition dipole moments (TDM) from the $4f^{13}6s^25d \: (J=2)$ state to the $4f^{13}6s^26p_{3/2} \: (J=3)$ and $4f^{13}6s^26p_{3/2} \: (J=2)$ states. The reduced matrix elements of the TDM for $4f^{13}6s^26p_{3/2} \: (J=3)$ and $4f^{13}6s^26p_{3/2} \: (J=2)$ states are 0.8587~a.u. and 0.3203~a.u., respectively, where a.u. denotes atomic unit.


\providecommand{\noopsort}[1]{}\providecommand{\singleletter}[1]{#1}%